\documentclass[
]{ceurart}

\usepackage{enumitem}            
\usepackage[numbers]{natbib}
\usepackage{amsmath}             
\usepackage{url}
\usepackage{hyperref}

\usepackage{booktabs}            
\usepackage{tabularx}            
\usepackage{longtable}           
\usepackage{array}               
\usepackage{adjustbox}           
\usepackage{pdflscape}           

\usepackage{graphicx}            
\usepackage{subcaption}          
\usepackage{caption}             
\usepackage{listings}            
\usepackage{xcolor}              
\usepackage{colortbl}
\usepackage{tcolorbox}
\tcbset{
  colback=white,
  colframe=blue!40!black,
  coltitle=blue!20!black,
  fonttitle=\bfseries,
  boxsep=2pt,
  arc=3pt
}

\definecolor{headerblue}{RGB}{220,230,255}
\definecolor{rowgray}{RGB}{245,245,245}

\usepackage{tikz}                
\usetikzlibrary{positioning, arrows.meta}  

\usepackage{url}

\usepackage{float}

\usepackage{listings}
\begin{document}

\copyrightyear{2026}
\copyrightclause{Copyright for this paper by its authors.
  Use permitted under Creative Commons License Attribution 4.0
  International (CC BY 4.0).}

\conference{Preprint. July 2026.}

\title{Graph Construction and Matching for Imperative Programs using Neural and Structural Methods}

\author[1]{Arshad Beg}[
    orcid=0009-0004-6939-0411,
    email=arshad.beg@mu.ie
]
\cormark[1]
\fnmark[1]

\author[1]{Diarmuid O'Donoghue}[
    orcid=0000-0002-3680-4217,
    email=diarmuid.odonoghue@mu.ie
]
\fnmark[1]

\author[1]{Rosemary Monahan}[
    orcid=0000-0003-3886-4675,
    email=rosemary.monahan@mu.ie
]
\fnmark[1]

\address[1]{Maynooth University, Co. Kildare, Ireland}

\cortext[1]{Corresponding author.}
\fntext[1]{These authors contributed equally.}

\begin{abstract}
  Reusing verification artefacts requires identifying structural and semantic similarities across programs and their specifications. In this paper, we focus on graph construction as a foundational step toward this goal. We present a pipeline that converts imperative programs and their annotations into typed, attributed graphs. Our experiments cover datasets including C with ACSL, Java with JML, and Dafny programs. The pipeline integrates abstract syntax tree parsing with semantic embeddings derived from models such as SentenceTransformer and CodeBERT. This enables the generation of graph representations that capture both structural relationships and semantic context. Our results show that consistent graph representations can be constructed across different languages and annotation styles. This work provides a practical basis for future steps in semantic enrichment and approximate graph matching for scalable verification artefact reuse.
\end{abstract}

\begin{keywords}
  Graph Construction \sep 
  Graph Matching \sep 
  Language Syntax Tree Parsing \sep
  Large Language Models
\end{keywords}

\maketitle

\section{Introduction}

Reusing verification artefacts such as specifications, contracts, and proofs remains a challenging task in software verification. Although large repositories of such artefacts exist, their reuse is still largely manual. Developers must search for relevant artefacts and adapt them to new contexts, even when similar solutions already exist. This process is difficult because artefacts often differ in syntax, abstraction level, and domain-specific vocabulary. In our long-term research vision \cite{Beg2026aLIFR}, we identified reuse as a key problem that requires principled mechanisms for discovering semantic correspondences between artefacts. Rather than generating new artefacts, the focus is on identifying structural and semantic similarities across existing ones. This can be understood as a problem of semantic matching under partial equivalence, where only fragments of behaviour or intent may align across different implementations.

To address this, we proposed representing verification artefacts as typed, attributed graphs \cite{Beg2026aLIFR}. In this representation, nodes capture semantic elements such as variables, predicates, transitions, and proof obligations, while edges encode relationships such as control-flow, dataflow, and logical dependencies. This abstraction allows programs, specifications, and proofs to be treated uniformly, enabling comparison across heterogeneous artefacts. However, graph structure alone is not sufficient. Important semantic information is also embedded in identifiers, comments, and logical expressions. To capture this, large language models (LLMs) can be used to generate embeddings that enrich graph nodes with semantic information. These embeddings allow the system to identify similarities that are not visible at the structural level.

\begin{figure}[t]
    \centering
    \includegraphics[width=\textwidth]{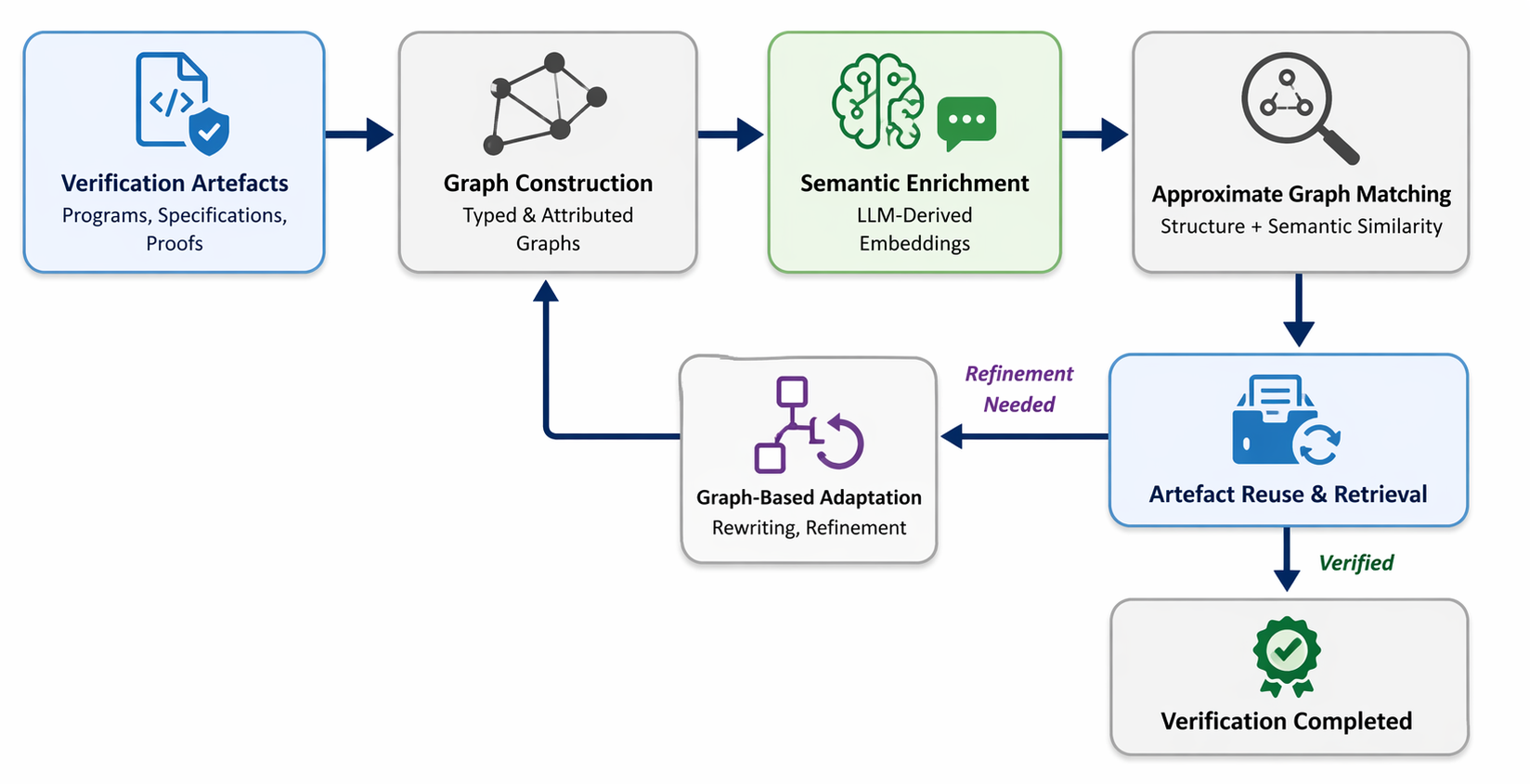}
    \caption{Workflow for verification artefact reuse via hybrid graph matching and LLM-based semantic enrichment.
    Verification artefacts are transformed into structured graphs, enriched with semantic embeddings derived from large language model (in our work, we used BERT model i.e. SentenceTransformer (all-MiniLM-L6-v2), CodeBERT and Graph Neural Netowrk(GNN)), and matched approximately using both structural and semantic similarity. Retrieved artefacts are either directly reused to complete verification or adapted through graph-based refinement.}
    \label{fig:llm-graph-reuse-workflow}
\end{figure}

In this paper, we focus on the first stage of this broader vision: graph construction for verification artefacts. We implement and evaluate a pipeline that translates imperative programs into typed, attributed graphs, forming the foundation for subsequent semantic enrichment and matching. The overall workflow, illustrated in Figure~\ref{fig:llm-graph-reuse-workflow}, outlines how graph construction integrates with LLM-based enrichment and approximate graph matching to support artefact reuse.

\paragraph{Key Contributions.}
We list down the key contributions:
\begin{itemize}
    \item We present a unified, end-to-end pipeline for graph construction and matching of imperative programs across multiple languages, including C, Java, and C\#, along with their specification frameworks (ACSL, JML, and Dafny).
    
    \item We design and implement automated, rule-based transformations to generate specification-augmented datasets (C--ACSL, Java--JML, and C\#--Dafny).
    
    \item We develop graph construction techniques that encode both program structure and formal specifications as typed, attributed graphs using AST-based and pattern-driven approaches.
    
    \item We integrate neural semantic representations using transformer-based models (SentenceTransformer and CodeBERT) to enrich graph nodes and enable similarity-based matching.
    
    \item We demonstrate the feasibility of cross-language artefact matching through embedding-based similarity, providing a foundation for scalable verification artefact reuse.
\end{itemize}

The structure of the paper is as follows: Section \ref{sec:syntactdiff} identifies the syntactical difference of the selected languages. Section \ref{sec:related_work} presents the related work. Section \ref{sec:methodology} comprises three subsections: the experimental setup, which explains dataset selection, virtual environment settings and the research workflow. Section \ref{sec:graph-construction} explains in detail the implementation of graph construction and matching for imperative programs. Section \ref{sec:conclusion} concludes the paper, with a focused subsection \ref{subsec:discussion} on lessons learned.

\section{Syntatical Differences:} 
\label{sec:syntactdiff}

There are syntactical differences between annotations in \textit{OpenJML}, \textit{ACSL}, and \textit{Dafny}, even though they often serve similar purposes (e.g., specifying preconditions, postconditions, and invariants). Below is a comparative overview of their syntactic styles.

\paragraph{OpenJML (Java + JML):}

OpenJML is based on Java and uses annotations embedded in special comments (e.g., \texttt{//\textbackslash @} or \texttt{/*\textbackslash @ \ldots @*/}). Specifications closely resemble Java syntax.

\begin{verbatim}
//@ requires x > 0;
//@ ensures \result == x + 1;
public int increment(int x) {
    return x + 1;
}
\end{verbatim}

It employs backslash-prefixed keywords such as \texttt{\textbackslash result}, \texttt{\textbackslash old}, \texttt{\textbackslash forall}, and \texttt{\textbackslash exists}.

\paragraph{ACSL (ANSI/ISO C Specification Language):}

ACSL is designed for annotating C programs. It uses block comments (\texttt{/*\textbackslash @ \ldots */}) and line comments (\texttt{//\textbackslash @}) for specifications. Its syntax is more mathematical, with some overlap with JML but adapted to C semantics.

\begin{verbatim}
/*@ requires x > 0;
    ensures \result == x + 1;
*/
int increment(int x) {
    return x + 1;
}
\end{verbatim}

Like JML, ACSL uses \texttt{\textbackslash result}, \texttt{\textbackslash old}, \texttt{\textbackslash forall}, and \texttt{\textbackslash exists}, but includes C-specific typing and memory modeling.

\paragraph{Dafny:}

Dafny is a programming language designed with built-in support for specification and verification. Specifications are integrated directly into the language, eliminating the need for annotation comments.

\begin{verbatim}
method increment(x: int) returns (r: int)
    requires x > 0
    ensures r == x + 1
{
    r := x + 1;
}
\end{verbatim}

It uses native keywords such as \texttt{requires}, \texttt{ensures}, \texttt{assert}, \texttt{assume}, and \texttt{invariant}, and supports tight integration with SMT solvers and functional-style proofs. Table 1 summarises the syntactical differences of the three notations. 

\begin{table}[h]
\centering
\begin{tabular}{lccc}
\hline
\textbf{Feature} & \textbf{OpenJML} & \textbf{ACSL} & \textbf{Dafny} \\
\hline
Annotation style & \texttt{//\textbackslash @}, \texttt{/*\textbackslash @ \ldots @*/} & \texttt{//\textbackslash @}, \texttt{/*\textbackslash @ \ldots */} & Native keywords \\
Result keyword & \texttt{\textbackslash result} & \texttt{\textbackslash result} & Variable in signature \\
Quantifiers & \texttt{\textbackslash forall}, \texttt{\textbackslash exists} & \texttt{\textbackslash forall}, \texttt{\textbackslash exists} & \texttt{forall}, \texttt{exists} \\
Old value & \texttt{\textbackslash old}$(\cdot)$ & \texttt{\textbackslash old}$(\cdot)$ & \texttt{old}$(\cdot)$ \\
Scope & Java-like class/method & C-like function/global & Dafny methods/functions \\
Spec location & Comments & Comments & Inline in code \\
\hline
\end{tabular}
\caption{Comparison of syntactical features in OpenJML, ACSL, and Dafny.}
\end{table}

\section{Related Work}
\label{sec:related_work}

In \cite{Beg2026aLIFR}, we presented a long term research agenda, named as Learning Infused Formal Reasoning (LIFR), having three dimensions: contract synthesis, verification artefact reuse and semantic foundations. Our paper addresses the second dimension. The workflow illustrated in Figure~\ref{fig:llm-graph-reuse-workflow} is a consolidated one, available in \cite{Beg2026aLIFR}. It outlines how graph construction integrates with LLM-based enrichment and approximate graph matching to support artefact reuse. The summary of research agenda is also represented in \cite{Beg2026bLIFR}.

Early work on graph-based representations of programs and specifications demonstrated the potential for reuse through structural similarity. The Aris system \cite{pitu2013aris} generated graph representations of source code combined with formal specifications, enabling the propagation of specifications to new implementations with similar structures \cite{odo2014creating}. This approach was further explored in \cite{aiyankovil2021}, where specification transfer was achieved within a single iteration for similar implementations. Graph-based similarity has also been applied beyond software artefacts. The Dr Inventor system \cite{odo2015stimulating} used graph representations to identify latent similarities between research publications, supporting the transfer of ideas across domains. However, these approaches primarily demonstrated feasibility and were not evaluated as large-scale, robust systems.

More recent work in knowledge-graph alignment and schema matching provides stronger empirical evidence for combining structural and semantic methods. Systems such as WES+pArtLink \cite{LippolisKMZJNH23} and RAG-augmented table-to-knowledge-graph alignment \cite{VandermoorteleS24} show that LLM guidance improves both coverage and precision when used alongside structural retrieval. The KG-RAG4SM system \cite{abs-2501-08686} further demonstrates that grounding LLM reasoning in retrieved graph structures reduces hallucination by constraining the search space. Similar ideas have been explored in software engineering and ontology pipelines. Code-graph–based retrieval augmented with LLMs \cite{AliNB24} shows that LLMs are effective when operating over structured representations. Likewise, ontology-driven natural-language–to–SPARQL systems \cite{TufekTJEBSH24} demonstrate how structured knowledge can guide LLM interpretation. Studies in cultural-heritage knowledge graphs \cite{vasic2025knowledge} also report that hybrid approaches combining symbolic structure with LLM-based semantics outperform purely symbolic or purely neural methods.

Graph transformation techniques are also relevant for artefact reuse. Once correspondences between artefacts are identified, adaptation is required to fit new contexts. Graph rewriting provides a formal mechanism for expressing such changes, including insertions, deletions, and refinements, while preserving semantic properties \cite{Strecker08GraphinProofAssitants}. These techniques offer a principled way to modify verification artefacts after matching. We can draw a conclusion from literature that LLMs within iterative reasoning pipelines is becoming common. Frameworks such as LangChain \cite{topsakal2023creating} and AutoGPT support retrieval-augmented generation and multi-step reasoning, often combining multiple LLMs. These systems highlight the growing importance of integrating LLMs with structured workflows, which aligns with the hybrid approach explored in this work.

\section{Methodology}
\label{sec:methodology}

\subsection{Experiment Setup}

\paragraph{\textbf{GitHub Repository:}} To make our work publicly accessible, the artifacts are available at our GitHub repository: \url{https://github.com/arshadbeg/Graphs-Construction-Matching-TechnicalReport}.

\paragraph{\textbf{Dataset:}} A benchmark comprising fifty-six (56) C programs, each demonstrating varying levels of structural complexity, was used without modification from a publicly available GitHub repository \footnote{\url{https://github.com/ggranberry/intent_dataset}} originally employed for prompt engineering in \cite{Granberry2025a}. Consistent with the original design, these programs were organized into three predefined categories: (i) correct implementations, (ii) implementations showing clear differences between function variants, and (iii) implementations containing subtle differences within their function bodies. Each of the fifty-six base programs was examined across all three categories, resulting in a total of 168 evaluated instances. This structured categorisation supports a systematic analysis of tool behaviour across a controlled range of semantic and syntactic variations. The required experiment setup and graph construction and matching is executed within a Linux environment running on Windows Subsystem for Linux (WSL 2), hosted on a Windows 11 system equipped with an Intel Core Ultra 5 125U processor (3.60 GHz) and 32 GB of RAM (31.5 GB usable) operating under a 64-bit architecture.

\paragraph{\textbf{Virtual Environment Settings:}} To support the graph-based analysis conducted in this work, we configured an isolated Python environment using \texttt{venv}. All dependencies were installed locally within the environment, ensuring isolation from system-level packages and consistent execution across experiments. The installed libraries reside in the \texttt{site-packages} directory of the virtual environment, comprising a diverse stack including program analysis tools (e.g., \texttt{clang}, \texttt{javalang}), graph processing libraries (e.g., \texttt{networkx}, \texttt{torch\_geometric}), and machine learning frameworks (e.g., \texttt{torch}, \texttt{transformers}, \texttt{scikit-learn}). Additional numerical and symbolic computation support is provided by packages such as \texttt{numpy}, \texttt{scipy}, and \texttt{sympy}. 

\subsection{Workflow}

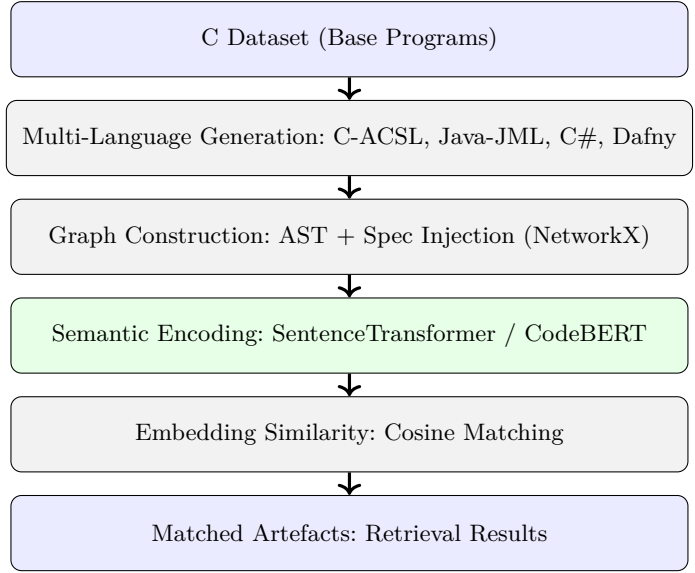
\begin{figure}[t]
\centering
\begin{tikzpicture}[
    node distance=0.3cm,
    every node/.style={
        draw,
        rounded corners,
        align=center,
        minimum height=1cm,
        minimum width=9cm,   
        inner sep=6pt,
        font=\small
    },
    data/.style={fill=blue!8},
    process/.style={fill=gray!10},
    model/.style={fill=green!10},
    arrow/.style={->, very thick}
]

\node[data] (cdata) {C Dataset (Base Programs)};
\node[process, below=of cdata] (gen) {Multi-Language Generation: C-ACSL, Java-JML, C\#, Dafny};
\node[process, below=of gen] (graph) {Graph Construction: AST + Spec Injection (NetworkX)};
\node[model, below=of graph] (embed) {Semantic Encoding: SentenceTransformer / CodeBERT};
\node[process, below=of embed] (match) {Embedding Similarity: Cosine Matching};
\node[data, below=of match] (output) {Matched Artefacts: Retrieval Results};

\draw[arrow] (cdata) -- (gen);
\draw[arrow] (gen) -- (graph);
\draw[arrow] (graph) -- (embed);
\draw[arrow] (embed) -- (match);
\draw[arrow] (match) -- (output);

\end{tikzpicture}

\caption{End-to-end workflow for graph construction and matching across C, Java, and Dafny pipelines with specification-aware transformations and neural embedding-based similarity.}
\label{fig:workflow}
\end{figure}

This work follows a multi-stage pipeline designed to support graph-based representation and matching of imperative programs across multiple languages and specification frameworks. The overall workflow begins with a unified C dataset, which serves as the base artefact collection. From this dataset, multiple language-specific and specification-augmented variants are systematically generated, including C with ACSL annotations, Java with JML specifications, C\# translations, and Dafny-augmented C\# programs. Each transformation stage is implemented using rule-based Python scripts that ensure structural consistency while adapting syntax and specification formats to the target language.
The overall workflow of the proposed approach is illustrated in Figure~\ref{fig:workflow}, which summarises the complete pipeline from dataset generation to graph-based matching. It highlights how C programs are systematically transformed into multiple specification-augmented representations, followed by graph construction, semantic embedding, and similarity-based artefact retrieval.

Once the multi-language dataset is constructed, each program is processed through a graph construction pipeline. This step converts source code into typed, attributed graphs using AST-based parsing techniques (or regex-guided structure extraction where full parsing is not available). For C, Java, and Dafny representations, control-flow constructs, specifications, and program statements are encoded as nodes and edges within directed graphs using \texttt{networkx}. In parallel, specification elements such as ACSL contracts, JML annotations, and Dafny invariants are explicitly injected into the graph structure, ensuring that formal semantics are preserved alongside program structure.

The next stage involves semantic enrichment and representation learning. Each constructed graph is linearised into textual form and encoded using transformer-based models such as SentenceTransformer (for C, Java, and C\# pipelines) and CodeBERT (for Dafny). These embeddings capture both syntactic patterns and latent semantic meaning across code and specifications. The resulting vectors are stored alongside graph structures to enable unified structural-semantic analysis.

Finally, pairwise similarity computation is performed using cosine similarity over embedding spaces to enable graph matching across artefacts. This allows identification of structurally and semantically similar programs across languages and specification levels. The complete workflow supports systematic artefact comparison, retrieval, and reuse, forming the basis for scalable verification-oriented program analysis.

\section{Graph Construction and Matching}
\label{sec:graph-construction}

\subsection{ACSL annotations generation for C Programs}

\begin{figure}[h]
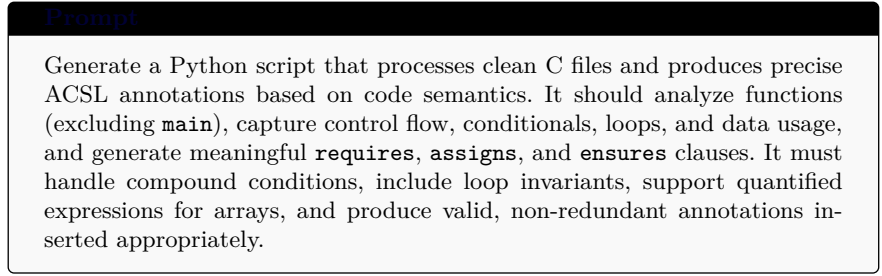

\centering
\begin{tcolorbox}[
    colback=gray!5,
    colframe=black,
    title=\textbf{Prompt},
    width=0.95\columnwidth,
    boxrule=0.5pt,
    arc=2pt
]
Generate a Python script that processes clean C files and produces precise ACSL annotations based on code semantics. It should analyze functions (excluding \texttt{main}), capture control flow, conditionals, loops, and data usage, and generate meaningful \texttt{requires}, \texttt{assigns}, and \texttt{ensures} clauses. It must handle compound conditions, include loop invariants, support quantified expressions for arrays, and produce valid, non-redundant annotations inserted appropriately.
\end{tcolorbox}
\caption{Prompt used to guide GPT-5.2 for ACSL annotation generation.}
\label{fig:prompt}
\end{figure}

Python script to automatically augment C programs with ACSL (ANSI/ISO C Specification Language) annotations is generated through GPT-5.2, guided by a prompt (see Fig.~\ref{fig:prompt}). The resulting script (267 lines) generates ACSL specifications from input C code. At the function level, conditional return patterns are identified, producing postconditions of the form:
\[
\texttt{ensures } (C) \Rightarrow \backslash result = E_1
\]
\[
\texttt{ensures } \neg(C) \Rightarrow \backslash result = E_2
\]
where $C$ is the branch condition and $E_1$, $E_2$ are returned expressions.

For loop constructs (\texttt{for}, \texttt{while}, \texttt{do}), heuristic rules infer loop indices, bounds, and array accesses. Preconditions ensure valid memory access and bounds, while invariants constrain index ranges and express array properties using quantified expressions. Termination is enforced via decreasing variants. Switch-case constructs are translated into conditional postconditions linking case labels to returned values. At the \texttt{main} function, runtime assertions are inserted after function calls based on simple argument properties, providing lightweight correctness checks.

\subsection{Graph Construction for C Programs with ACSL annotations}

The graph construction and matching pipeline is implemented in Python and follows a modular, end-to-end design. It integrates standard libraries such as \texttt{os} and \texttt{json} for file handling and data storage, along with \texttt{networkx} for graph representation and \texttt{numpy} for numerical computation. For parsing source code, the pipeline uses \texttt{clang.cindex}, configured with \texttt{libclang}, to extract abstract syntax trees (ASTs) from C programs. In addition, semantic representations are generated using the \texttt{SentenceTransformer} library with the \texttt{all-MiniLM-L6-v2} model. The pipeline is structured into distinct stages: graph construction, embedding generation, and similarity-based matching, ensuring a clear separation between structural and semantic processing.

Graph construction is performed by traversing the AST produced by Clang. Each node in the AST is converted into a node in a directed graph using \texttt{networkx.DiGraph}, where node attributes capture the type and textual content of the original AST element. Unique identifiers are generated using node metadata such as kind, line, and column information. Parent-child relationships in the AST are preserved as directed edges labeled \texttt{AST\_CHILD}, ensuring that the hierarchical program structure is retained. The resulting graphs are serialised in GraphML format using \texttt{networkx.write\_graphml}, allowing interoperability and further analysis. This stage captures the structural backbone of programs in a uniform and language-aware representation.

\paragraph{Generated GraphML Files:} The generated GraphML representation encodes the program structure as a directed, labelled graph in which each node corresponds to a syntactic construct derived from the program’s abstract syntax tree (AST). Each \texttt{<node>} element is uniquely identified by an \texttt{id} attribute that typically combines the node type, source location (e.g., line and column), and an internal identifier to ensure uniqueness. The semantic role of a node is captured \texttt{<data>} elements: the key \texttt{d0} denotes the syntactic category (e.g., \texttt{DECL\_STMT}, \texttt{VAR\_DECL}, \texttt{BINARY\_OPERATOR}), while \texttt{d1} stores an associated label such as a variable name (e.g., \texttt{i}, \texttt{low}) or repeats the operator/type when no explicit identifier exists. For instance, a declaration statement node (\texttt{DECL\_STMT}) is linked to a variable declaration node (\texttt{VAR\_DECL}) representing \texttt{i}, while expression nodes such as \texttt{BINARY\_OPERATOR} and \texttt{DECL\_REF\_EXPR} capture the structure of expressions involving program variables. This hierarchical organisation, complemented by edges (not shown here) representing parent–child relationships, enables the graph to faithfully model both control and data-relevant syntactic constructs of the original program.

\begin{figure}[t]
    \centering
    \includegraphics[width=\textwidth]{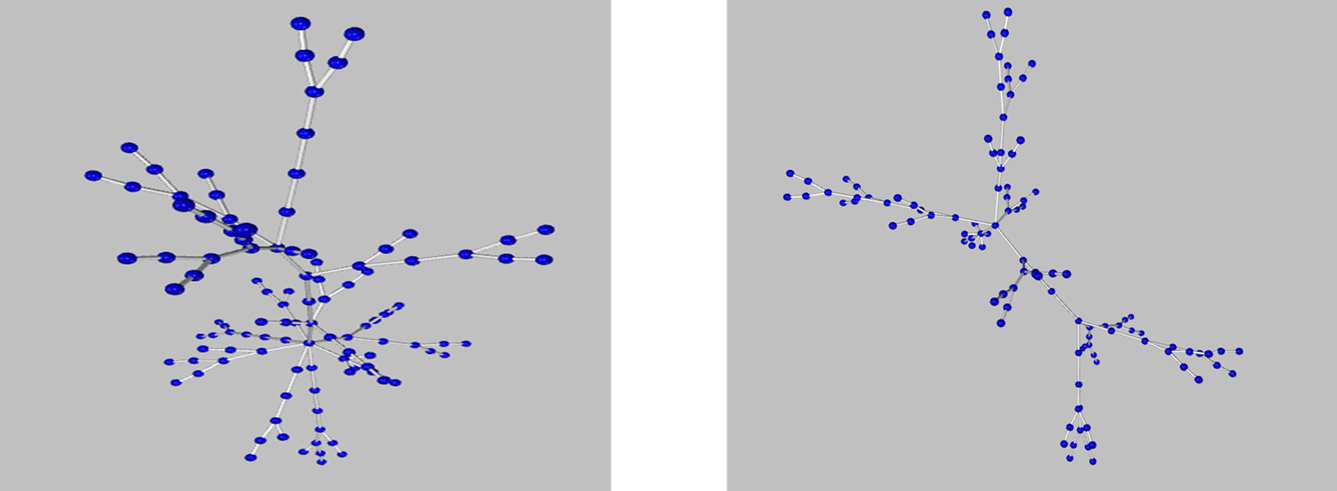}
    \caption{Graphs constructed for ACSL annotated C Programs of quicksort and merge}
    \label{fig:quick-merge-acsl}
\end{figure}

Semantic enrichment is achieved by transforming graph content into textual form and generating embeddings. Node-level textual attributes are aggregated into a single sequence using a simple traversal strategy. This text is then passed to the \texttt{SentenceTransformer} model to produce dense vector embeddings. A fallback mechanism ensures stability by assigning a zero vector when no textual content is available. The embeddings are converted into standard Python float representations to avoid datatype inconsistencies and stored in JSON format. This step introduces semantic information that complements the structural graph, enabling comparisons beyond syntactic similarity. Figure \ref{fig:quick-merge-acsl} show the graph constructed for quicksort and merge programs.

Graph matching is performed using pairwise similarity computation over the generated embeddings. The pipeline applies cosine similarity, implemented with \texttt{numpy}, to measure the closeness between embedding vectors. All program pairs are compared in a systematic manner, and similarity scores are recorded alongside file paths. The results are stored as structured JSON data, enabling downstream retrieval and analysis. The overall pipeline orchestrates these steps through directory traversal using \texttt{os.walk}, ensuring scalability across datasets. This combination of AST-based graph construction, transformer-based embeddings, and numerical similarity provides a practical foundation for approximate graph matching in verification artefact reuse.  

\paragraph{Discussion on output matches.JSON file}

The similarity results captured in the matches.json file represent a comprehensive pairwise graph-matching output where each C source file from the ACSL dataset is compared against every other file to quantify structural and semantic resemblance. Each JSON entry consists of two file paths (file1 and file2) and a corresponding cosine-based similarity score derived from their learned graph embeddings. The values span a broad spectrum, typically ranging from low similarity (0.35–0.60), indicating weak structural correspondence between unrelated algorithms (e.g., search versus cryptographic or hashing routines), to moderate similarity (0.60–0.80), where programs share partial algorithmic patterns such as loops, conditional logic, or common data-handling constructs. At the higher end (0.80–0.95), the matched pairs reflect strong semantic alignment, often occurring between variants of the same algorithm or closely related implementations such as sorting, searching, or numerical utility functions. These high-scoring pairs indicate that the graph representations successfully capture deeper syntactic and structural features beyond surface-level token similarity.

A particularly important aspect of this output format is its global comparability, where a single reference file (e.g., binarysearch.c) is evaluated against all other programs in the dataset, producing a ranked similarity distribution that reflects both near-duplicates and functionally related implementations. This dense comparison structure enables fine-grained clustering of programs into logical families, such as “basic,” “famous,” “mirror,” and “unique” categories, as well as cross-category relationships where algorithmic overlap exists despite differing contexts. The JSON structure thus serves not only as a storage format but also as an analytical artifact for downstream tasks such as plagiarism detection, code retrieval, and semantic clustering.

Finally, the distribution of similarity scores highlights the discriminative capability of the underlying graph embedding model: low scores consistently correspond to semantically divergent programs, while high scores correctly surface algorithmic equivalence even across stylistically different implementations. Intermediate values capture partial structural overlap, such as shared control-flow constructs or similar loop nesting patterns, which are common in algorithmic variations like sorting, searching, or array manipulation. Collectively, this similarity matrix forms a weighted relational map over the dataset, enabling both qualitative inspection of program relationships and quantitative evaluation of structural similarity across the ACSL corpus.

\subsection{Java Files Dataset}

The Java files dataset used in our experiments is systematically generated from the original C dataset. This ensures consistency across languages while preserving the structural characteristics of the source programs. The transformation is implemented as a Python-based pipeline that processes each C file and produces a corresponding Java version. The script relies on standard libraries such as \texttt{os} for directory traversal and file handling, and \texttt{re} for regular expression–based pattern matching. As an initial step, the pipeline removes C-specific preprocessor directives (e.g., \texttt{\#include}, \texttt{\#define}) to ensure compatibility with Java syntax. This is followed by type normalisation, where C types such as pointers to \texttt{char} are mapped to Java \texttt{String}, while primitive types are preserved where possible.

The core transformation operates through a sequence of rule-based conversions applied line by line. Input/output constructs are adapted by converting \texttt{printf} statements into \texttt{System.out.println} calls and \texttt{scanf} statements into \texttt{Scanner}-based input handling. Function signatures are rewritten to match Java conventions, including transforming the \texttt{main} function into \texttt{public static void main(String[] args)}. Additionally, C \texttt{struct} definitions are translated into static inner Java classes, with fields extracted and converted using the same type-mapping rules. Function definitions are converted into static Java methods by parsing return types and parameters, ensuring that method signatures remain consistent with the original program logic.

Finally, each transformed program is wrapped within a Java class, with necessary imports such as \texttt{java.util.*} and a shared \texttt{Scanner} instance for input handling. The script preserves the relative directory structure of the original dataset using \texttt{os.path.relpath}, enabling traceability between C and Java versions. Output files are generated with sanitised class names to ensure validity in Java. The entire dataset is processed recursively using \texttt{os.walk}, allowing scalable conversion across large collections of programs. This automated transformation provides a consistent multi-language dataset, enabling comparative analysis and supporting downstream tasks such as graph construction and matching.

\subsection{JML Generation}

The JML generation pipeline operates on the Java dataset produced earlier and automatically augments each file with formal specifications. It is implemented in Python using standard libraries such as \texttt{os} for directory traversal and file management, \texttt{re} for pattern-based parsing, and \texttt{typing} for structured data handling. The pipeline begins by identifying valid Java files through simple filtering rules that exclude any residual C/C++ preprocessing directives. Additional cleaning ensures that such directives are removed if present, maintaining syntactic correctness. File discovery is handled recursively using \texttt{os.walk}, and utility functions manage reading and writing while preserving the directory structure of the dataset.

The core of the approach relies on regular expressions to extract structural elements from Java code. Patterns are defined for detecting classes, methods, and fields, enabling lightweight static analysis without requiring a full parser. Field extraction is used to generate class-level invariants, where heuristic rules assign constraints based on data types. For example, numeric fields are constrained to be non-negative, boolean fields are restricted to valid truth values, and reference types are required to be non-null. These invariants are inserted immediately after class declarations, ensuring that global properties of the program state are explicitly captured.

Method-level specifications are generated by analysing parameter lists and return types. For each method, \texttt{requires} clauses enforce non-null conditions on parameters, while \texttt{ensures} clauses specify basic guarantees about return values when applicable. An \texttt{assignable} clause is also included to indicate permissible state modifications. These contracts are inserted directly above method declarations, preserving readability and structure. The overall transformation is applied consistently across all files, producing a JML-annotated dataset that retains the original program logic while introducing lightweight formal specifications suitable for further analysis and verification tasks.

\subsection{Graph Construction and Matching for JML augmented Java Files}

\begin{figure}[t]
    \centering
    \includegraphics[width=\textwidth]{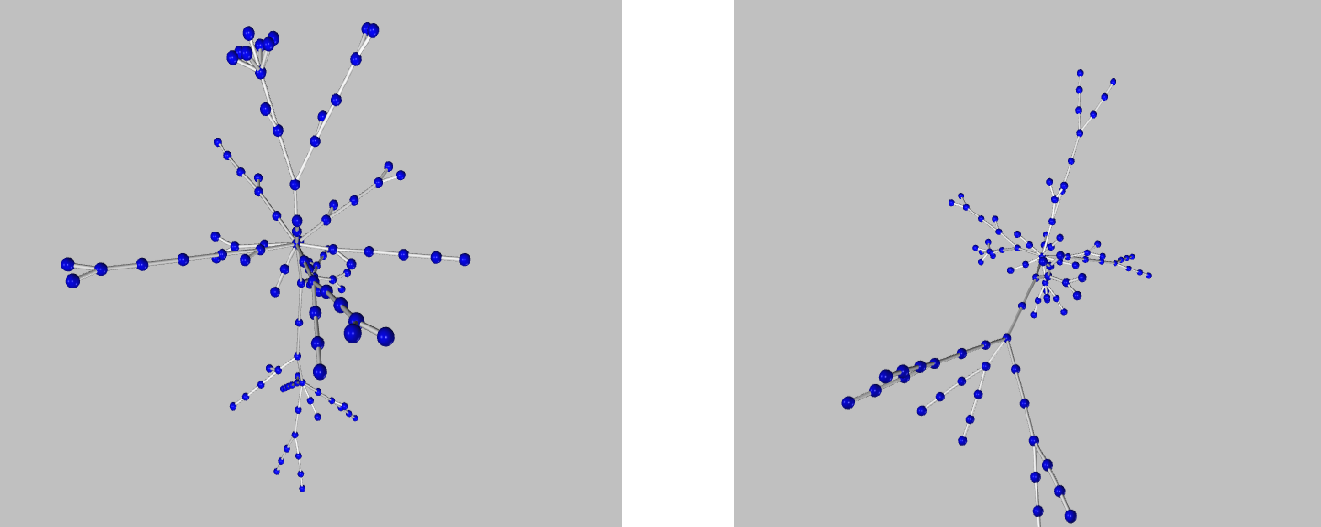}
    \caption{Graphs constructed for JML annotated Java Programs of quicksort and merge}
    \label{fig:quick-merge-jml}
\end{figure}

The graph construction process for JML-augmented Java files is tailored to capture both program structure and embedded specifications in a unified representation. Instead of relying on compiler bindings, the pipeline uses the \texttt{javalang} library to parse Java source code into an abstract syntax tree. This choice enables lightweight and language-specific parsing directly within Python. The AST is traversed recursively, and each syntactic element is mapped into a node within a directed graph using \texttt{networkx}. Node attributes store both the structural type (e.g., class, method, expression) and a textual label derived from identifiers or values. A distinctive aspect of this pipeline is the explicit extraction of JML annotations from comments using regular expressions. Since JML specifications are not part of the standard AST, they are injected as separate nodes and linked to the file-level node via dedicated edges. This design ensures that formal specifications are preserved and made available alongside structural program elements within the same graph. Figure \ref{fig:quick-merge-jml} show the graph constructed for quicksort and merge programs.

For matching, the pipeline combines this enriched graph representation with semantic encoding techniques. All node-level textual content, including JML annotations, is aggregated and transformed into dense vector embeddings using the \texttt{SentenceTransformer} model. These embeddings provide a compact representation of both code semantics and specification intent. Similarity between Java artefacts is then computed using cosine similarity implemented with \texttt{numpy}, allowing efficient pairwise comparison across the dataset. The resulting similarity scores reflect not only structural resemblance but also alignment in specification-level intent, which is particularly important for JML-annotated programs. By integrating extracted specifications directly into the graph and embedding space, the pipeline supports a more expressive form of matching that goes beyond pure code structure.  

\paragraph{Discussion on output matches.JSON file}

The Java–JML graph matching results are stored in a structured JSON format where each entry corresponds to a pairwise comparison between two source files along with their computed similarity score. Each record contains the fields file1, file2, and a normalised similarity value in the range [0, 1], capturing the degree of structural and behavioral alignment between Java implementations annotated with JML specifications. For instance, comparisons involving duplicate.java against a range of basic algorithmic programs such as sorting, searching, and numerical utilities show moderate to high similarity values, typically ranging from approximately 0.66 to 0.88, reflecting shared control-flow patterns and reusable logic structures. Higher similarity scores (above 0.85) are observed for closely related transformations or near-identical implementations, while lower values (around 0.60–0.70) indicate weaker structural correspondence, often due to differing algorithmic strategies or specification constraints introduced by JML annotations.

Across the broader dataset, spanning correct, obvious, and subtle variants, as well as categories like mirror, famous, and unique, the similarity distribution remains diverse but consistent with expected semantic proximity. Near-identical program pairs, such as intentionally duplicated or minimally altered variants, achieve scores close to 1.0, whereas algorithmically distinct programs such as search versus cryptographic or hashing routines fall toward the lower bound of the spectrum. This graded distribution demonstrates that the matching process effectively captures both syntactic and specification-driven similarities introduced through JML, while also distinguishing semantically different implementations even when they share superficial structural patterns.

\subsection{CSharp Files Generation}

The generation of C\# files from the original C dataset is performed using a Python rule-based transformation pipeline that maps C constructs to their closest C\# equivalents while preserving program structure. C headers such as \texttt{\#include <stdio.h>} are replaced with \texttt{using System;}, \texttt{printf} is converted to \texttt{Console.WriteLine}, and \texttt{scanf} is translated into appropriate \texttt{Console.ReadLine()} parsing expressions according to data type. The \texttt{main} function becomes \texttt{static void Main(string[] args)}, C data types are mapped to C\# equivalents, including pointer-to-string conversions, and \texttt{struct} definitions are rewritten as C\# \texttt{struct} blocks. Function definitions are converted into static methods with reconstructed parameter lists, while loop constructs are largely preserved because of their syntactic similarity. The transformed code is wrapped with the required boilerplate, written as corresponding \texttt{.cs} files, and stored using the original directory hierarchy, producing a consistent multi-language dataset aligned with the source C programs.

\subsection{Dafny Generation}

Dafny programs are generated from the C\# dataset using a Python transformation that extracts behavioural structure and expresses it in a verification-oriented form. Rather than full parsing, the approach uses carefully designed pattern matching to identify method definitions, control-flow constructs, return statements, and method bodies. Parameters and return types are simplified into Dafny-compatible forms, preserving key computational behaviour while avoiding unnecessary source-language syntax.

The transformation emphasises logical intent over syntactic fidelity. Conditional return patterns are converted into \texttt{ensures} clauses, with dual postconditions generated when both \texttt{if} and \texttt{else} branches return values. Loop constructs are analysed to infer basic invariants and termination measures, including iteration bounds and decreasing variants required for verification. The resulting standalone Dafny methods preserve the control-flow semantics of the original C\# programs while embedding formal specifications. Generated files retain the original directory structure to support traceability and verification workflows.

\subsection{Graph Construction and Matching from Dafny augmented CSharp Files}

Graph construction represents both executable structure and formal specifications within a unified graph. Implemented in Python with \texttt{networkx}, the pipeline identifies Dafny constructs such as methods, functions, and specifications using regular expressions. Each file forms a root node connected to method nodes labeled with method names and parameters. Method bodies are extracted and processed systematically, while specification elements including \texttt{requires}, \texttt{ensures}, invariants, and assertions are represented as dedicated nodes linked to their corresponding methods. This preserves both program logic and verification conditions in a single graph.

Method bodies are further decomposed into statement and control-flow nodes representing constructs such as \texttt{if}, \texttt{while}, and return expressions. Loop invariants and assertions are attached as child nodes, while unique identifiers and edges labeled \texttt{AST\_CHILD} maintain consistent parent-child relationships. The resulting graphs are serialised in GraphML format for downstream analysis.

For matching, node text is aggregated and encoded using CodeBERT. The tokenised representation is processed to obtain contextual embeddings, with the \texttt{[CLS]} token serving as the graph representation. Pairwise cosine similarity is then computed using \texttt{numpy}, producing similarity scores that capture both code semantics and specification intent for artifact retrieval and alignment.

\paragraph{Discussion on output matches.JSON file}

We summarise similarity results for Dafny implementations anchored on quicksort.dfy, which consistently exhibits very high structural similarity across the dataset. Most comparisons across correct, obvious, subtle, mirror, unique, and famous variants cluster in the 0.95--1.00 range, indicating strong semantic and syntactic overlap despite naming or structural differences. High similarity with programs such as reverse, rotate, sorted, kmp, levenshtein, and voting suggests shared algorithmic patterns and reusable verification scaffolding. Mirror and unique transformations preserve logical structure while renaming or restructuring control flow, maintaining high similarity. A notable outlier is insert.dfy with near-zero similarity, reflecting a fundamentally different algorithmic structure. Overall, quicksort acts as a hub implementation, with most files exhibiting near-duplicate verified logic and consistent specification-driven design across verified and experimental subsets.

\section{Conclusion}
\label{sec:conclusion}

This paper presented a unified framework for graph construction and matching across imperative programs and formal specifications, covering C with ACSL, Java with JML, and Dafny. Automated pipelines transform source programs and annotations into typed, attributed graphs that preserve control flow, data usage, and specification constructs, enabling a common representation across heterogeneous artifacts. The results demonstrate that comparable graph structures can be systematically constructed while retaining both syntactic and semantic information.

A key contribution is the integration of specification-aware graph construction with embedding-based representations. By incorporating ACSL contracts, JML specifications, and Dafny invariants into graph structures and combining them with SentenceTransformer and CodeBERT embeddings, the approach captures both program behaviour and intended correctness properties. This enables similarity assessment beyond surface syntax by jointly considering structural alignment and specification intent.

We can say that the proposed framework provides a practical basis for scalable verification artifact reuse through graph representations. Future work will investigate approximate graph matching, semantic alignment, and cross-language artifact adaptation by combining structural constraints with learned representations, advancing automated and reusable verification workflows.

\subsection{Discussion}
\label{subsec:discussion}

While the proposed pipeline demonstrates the feasibility of constructing and matching graph-based representations across multiple programming languages and specification frameworks, several limitations remain. First, the transformation stages rely heavily on rule-based and regex-driven extraction rather than full semantic parsing. This design choice improves scalability and simplicity but can introduce brittleness when encountering complex or non-standard code patterns. As a result, certain edge cases in control flow or deeply nested constructs may not be accurately captured in the graph representation. Second, the quality of semantic matching is strongly dependent on the embedding models used (SentenceTransformer and CodeBERT). Although these models provide useful general-purpose representations, they are not specifically trained for formal verification artefacts or specification-heavy code. This can lead to reduced sensitivity in distinguishing fine-grained semantic differences, particularly when two programs share similar structure but differ in specification intent. Additionally, cosine similarity over global embeddings may oversimplify richer structural relationships encoded in graphs.

From a practical perspective, we also observed that specification injection (ACSL, JML, and Dafny) improves expressiveness but introduces heterogeneity in representation density across datasets. This affects comparability and may bias matching toward specification-rich files. Furthermore, the pipeline assumes consistent formatting across generated datasets, which may not always hold in real-world codebases. These observations highlight the need for more robust parsing mechanisms and domain-adapted embedding models. Despite these limitations, several important lessons were learned. A key insight is that separating structural graph construction from semantic embedding significantly improves modularity and extensibility of the pipeline. Another lesson is that even lightweight, rule-based specification extraction can substantially enhance downstream matching performance when combined with neural representations. Finally, the experiments reinforce the importance of integrating symbolic structure (graphs) with learned representations, as neither approach alone is sufficient for reliable cross-language artefact matching. These findings provide a foundation for future work on more precise, semantics-aware graph learning frameworks for program verification artefacts.

\section*{Use of AI-Assisted Tools}

We acknowledge the use of free version of GPT-5.2 for refining the textual presentation of the paper.
The model was applied to improve clarity and coherence. Followed by this, the text has been thoroughly
reviewed and discussed by all authors to ensure accuracy and integrity.

\section*{Acknowledgement}
This work has been supported by the European Commission as part of the ADAPT Centre for Digital Content Technology which is funded under the SFI Research Centres Programme (Grant 13/RC/2106) and is co-funded under the European Regional Development Fund.

\bibliographystyle{plain}
\bibliography{ftc2026Bib}

\end{document}